\documentclass[12pt,reqno]{amsart}

\usepackage{epsfig}
\usepackage{amsmath, amsthm, amsfonts}
\usepackage{graphicx}

\numberwithin{equation}{section}

\newcommand{\di}{\displaystyle}

\newcommand{\R}{{\mathbb R}}
\newcommand{\C}{{\mathbb C}}
\newcommand{\N}{{\mathbb N}}

\renewcommand{\Re}{{\operatorname{Re\,}}}
\renewcommand{\Im}{{\operatorname{Im\,}}}

\newcommand{\Tr}{{{\operatorname{Tr}}}}

\newcommand{\dist}{{\operatorname{dist}}}

\newcommand{\al}{\alpha}
\newcommand{\be}{\beta}
\newcommand{\ga}{\gamma}
\newcommand{\Ga}{\Gamma}

\newcommand{\ep}{\varepsilon}
\newcommand{\de}{\delta}
\newcommand{\De}{\Delta}

\newcommand{\sg}{\sigma}

\newcommand{\om}{\omega}

\newtheorem{theo}{{\sc \bf Theorem}}[section]

\begin{document}

\title[Exact solution of the six-vertex model]
{Exact solution of the six-vertex model with domain wall boundary conditions.
Critical line between ferroelectric and disordered phases}

\author{Pavel Bleher}
\address{Department of Mathematical Sciences,
Indiana University-Purdue University Indianapolis,
402 N. Blackford St., Indianapolis, IN 46202, U.S.A.}
\email{bleher@math.iupui.edu}

\author{Karl Liechty}
\address{Department of Mathematical Sciences,
Indiana University-Purdue University Indianapolis,
402 N. Blackford St., Indianapolis, IN 46202, U.S.A.}
\email{kliechty@math.iupui.edu}

\thanks{The first author is supported in part
by the National Science Foundation (NSF) Grant DMS-0652005.}

\date{\today}

\begin{abstract} This is a continuation of the papers \cite{BF} of Bleher and Fokin
and \cite {BL} of Bleher and Liechty,
in which the large $n$ asymptotics is obtained for the partition function $Z_n$ of the
six-vertex model with domain wall boundary conditions in the disordered and
ferroelectric phases, respectively. In the present paper
we obtain the large $n$ asymptotics of $Z_n$ on the critical line between these two phases. 
\end{abstract}

\maketitle

\section{Introduction and formulation of the main result}

\subsection{Definition of the model}
The six-vertex model, or the model of two-dimensional ice, is stated on a square $n\times n$
lattice with arrows on edges. The arrows obey the rule that at every vertex there 
are two arrows 
pointing in and two arrows pointing out. Such rule is sometimes 
called the {\it ice-rule}. There are only six possible configurations of arrows at each 
vertex, hence the name of the model, see Fig.~1. 

%%%%%%%%%%%% Fig.  %%%%%%%%%%%%%%
\begin{center}
 \begin{figure}[h]\label{arrows}
\begin{center}
   \scalebox{0.52}{\includegraphics{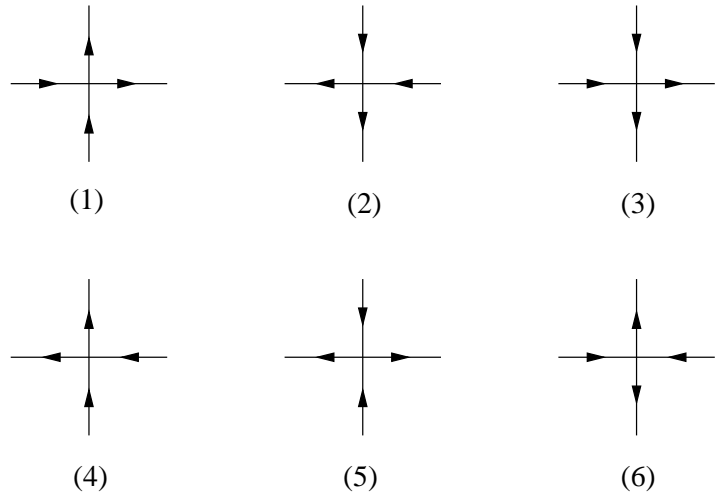}}
\end{center}
        \caption[The six arrow configurations allowed at a vertex]{The six arrow configurations allowed at a vertex.}
    \end{figure}
\end{center}
%%%%%%%%%%%%%%%%%%%%%%%%%%%%%%%%%%

We will consider the {\it domain wall boundary conditions} (DWBC), 
in which the arrows on the upper and lower boundaries point in the square, 
and the ones on the left and right boundaries point out. 
One possible configuration with DWBC on the $4\times 4$ lattice is shown on Fig.~2.
%%%%%%%%%%%% Fig.  %%%%%%%%%%%%%%
\begin{center}
 \begin{figure}[h]\label{DWBC}
\begin{center}
   \scalebox{0.52}{\includegraphics{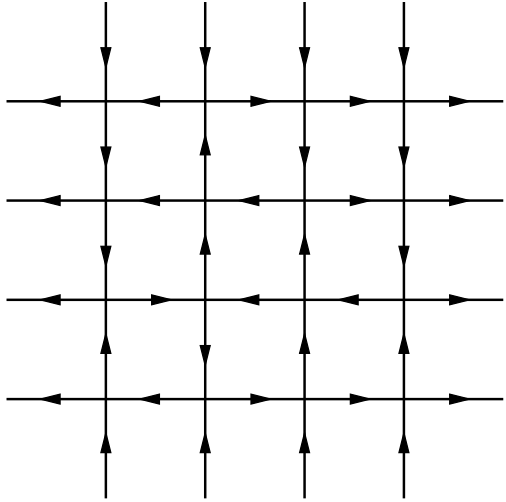}}
\end{center}
        \caption[An example of $4\times4$ configuration]
{An example of $4\times4$ configuration with DWBC.}
    \end{figure}
\end{center}
%%%%%%%%%%%%%%%%%%%%%%%%%%%%%%%%%%

For each possible vertex state we assign a weight $w_i,\; i=1,\dots,6$, 
and define, as usual, the partition function, as a sum over all possible 
arrow configurations of the product of the vertex weights,
\begin{equation}\label{lattice_11}
Z_n=\sum_{{\rm arrow\; configurations}\;\sigma}w(\sigma),
\qquad w(\sigma)=\prod_{x\in V_n} w_{\sg(x)}=\prod_{i=1}^6w_i^{N_i(\sigma)},
\end{equation}
where $V_n$ is the $n\times n$ set of vertices,
$\sg(x)\in\{1,\ldots,6\}$ is the vertex state of the configuration $\sg$
at the vertex $x$, according to Figure 1, and
$N_i(\sigma)$ is the number of vertices of the vertex state $i$ in  
the configuration $\sg$. The sum is taken over all possible configurations
obeying the given boundary condition. The Gibbs measure is defined then
as
\begin{equation}\label{lattice_12}
\mu_n(\sg)=\frac{w(\sg)}{Z_n}\,.
\end{equation}
Our main goal is to obtain the large $n$ asymptotics of the partition function $Z_n$.

 In general, the six-vertex model has six parameters: the weights $w_i$.  
However, by using some conservation laws we can reduce these to only two parameters.  
Namely, first we reduce to the case 
\begin{equation}\label{par}
w_1=w_2\equiv a, \quad
w_3=w_4\equiv b, \quad
w_5=w_6\equiv c,
 \end{equation}
and then, by using the identity,
\begin{equation}\label{cl_17}
Z_n(a,a,b,b,c,c)=c^{n^2}Z_n\left(\frac{a}{c},\frac{a}{c},\frac{b}{c},\frac{b}{c},1,1\right),
\end{equation}
to the two parameters, $\frac{a}{c}$ and $\frac{b}{c}\,$.
For details on how we make this reduction, see, e.g., the works \cite{AR} of Allison and Reshetikhin,
\cite{FS} of Ferrari and Spohn, or the work \cite{BL}. 

\subsection {The phase diagram} Introduce the parameter
\begin{equation}\label{pf1}
\Delta=\frac{a^2+b^2-c^2}{2ab}\,.
\end{equation}
The phase diagram of the six-vertex model consists of the following three
 regions: the ferroelectric phase region, $\Delta > 1$; the anti-ferroelectric phase region, 
$\Delta<-1$; and, the disordered phase region, $-1<\Delta<1$, see, e.g., \cite{LW}. 
In these three regions we parameterize the weights in the standard way:
in the ferroelectric phase region,
\begin{equation}\label{pf4}
a=\sinh(t-\ga), \quad
b=\sinh(t+\ga), \quad
c=\sinh(2|\ga|), \quad
0<|\ga|<t,
\end{equation}
in the anti-ferroelectric phase region,
\begin{equation}\label{pf5}
a=\sinh(\ga-t), \quad
b=\sinh(\ga+t), \quad
c=\sinh(2\ga), \quad
|t|<\ga,
\end{equation}
and in the disordered phase region,
\begin{equation}\label{pf6}
a=\sin(\ga-t), \quad
b=\sin(\ga+t), \quad
c=\sin(2\ga), \quad
|t|<\ga.
\end{equation}
The phase diagram of the six-vertex model is shown in Fig.~3.
\begin{center}
 \begin{figure}[h]\label{PhaseDiagram1}
\begin{center}
   \scalebox{0.5}{\includegraphics{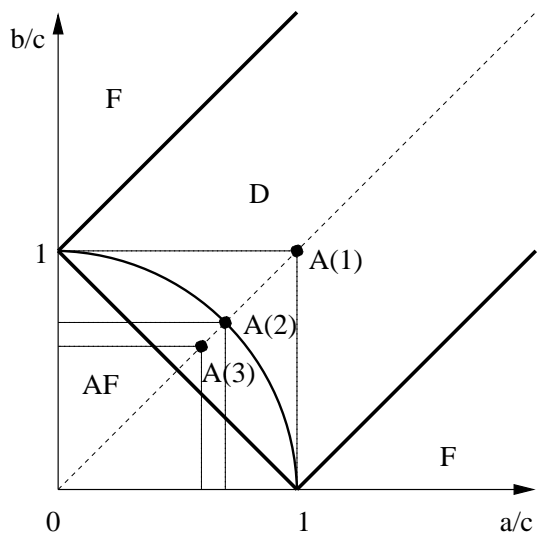}}
\end{center}
        \caption[The phase diagram of the model]{The phase diagram of the model, 
where {\bf F}, {\bf AF} and {\bf D} mark ferroelectric, antiferroelectric,  
and disordered  phase regions, respectively. The circular arc corresponds to 
the so-called "free fermion" line, when $\Delta=0$, and the three
dots correspond to 1-, 2-, and 3-enumeration of alternating sign matrices.}
    \end{figure}
\end{center}
%%%%%%%%%%%%%%%%%%%%%%%%%%%%%%%%%%
The phase diagram  and the Bethe-Ansatz solution  of the six-vertex model for 
periodic and anti-periodic boundary conditions are thoroughly
discussed in the works of Lieb \cite{Lieb1}-\cite{Lieb4}, Lieb, Wu \cite{LW},
Sutherland \cite{Sut}, Baxter \cite{Bax}, Batchelor, Baxter, O'Rourke, Yung \cite{BBOY}.
See also the work of Wu, Lin \cite{WL}, in which  the Pfaffian solution for the six-vertex
model with periodic boundary conditions is obtained on the free fermion line, $\Delta=0$.

As concerns the six-vertex model with DWBC, it is noticed by Kuperberg \cite{Kup}, 
that on the diagonal, 
\begin{equation}\label{pf2}
\frac{a}{c}=\frac{b}{c}=x,
\end{equation}
 the six-vertex model with DWBC is equivalent to the $s$-enumeration of alternating
sign matrices (ASM), in which the weight of each such matrix is equal to $s^{N_-}$, 
where $N_-$ is the number of $(-1)$'s in the matrix and $s=\frac{1}{x^2}$. The exact
solution for a finite $n$ is known for 1-, 2-, and 3-enumerations of ASMs, see the 
works by Kuperberg \cite {Kup} and Colomo and Pronko \cite{CP1} for a solution
based on the Izergin-Korepin formula. A fascinating story of the discovery 
of the ASM formula is presented in the book  \cite{Bre} of Bressoud.
On the free fermion line, $\ga=\frac{\pi}{4}$, the partition function of
the six-vertex model with DWBC has a very simple form: $Z_n=1$. For a nice
short proof of this formula see the work \cite{CP1} of Colomo and Pronko. 

Here we will discuss  the ferroelectric and disordered phase regions, and we will use 
parameterizations (\ref{pf4})and (\ref{pf6}). The ferroelectric phase region consists 
of two connected components. For the sake of concreteness we will assume that
\begin{equation}\label{pf6a}
\ga>0,
\end{equation}
which corresponds to the component where 
\begin{equation}\label{pf6b}
b>a+c.
\end{equation}
The parameter $\Delta$ in the ferroelectric phase region reduces to
\begin{equation}\label{pf6c}
\Delta=\cosh(2\ga).
\end{equation}

\subsection {Exact solution of $Z_n$ for finite $n$ in ferroelectric phase}
 The six-vertex model with DWBC was first introduced by Korepin in \cite{Kor}.  
In this paper, he derived an important recursion relation for $Z_n$, 
which was subsequently used by Izergin \cite{Ize} in deriving a determinantal 
formula for $Z_n$ in this model.  A detailed proof of this formula and its 
generalizations are given in the paper of Izergin, Coker, and Korepin \cite{ICK}, 
see also the papers of Korepin, Zinn-Justin \cite{KZ} 
and Kuperberg \cite {Kup} and the book of Bressoud \cite{Bre}.  
When the weights are parameterized according to (\ref{pf4}), the formula of Izergin-Korepin is
\begin{equation} \label{pf7}
Z_n=\frac{[\sinh(t-\ga)\sinh(t+\ga)]^{n^2}}{\left(
\prod_{j=0}^{n-1}j!\right)^2}\,\tau_n\,,
\end{equation}
where $\tau_n$ is the Hankel determinant,
\begin{equation} \label{pf8}
\tau_n=\det\left(\frac{d^{j+k-2}\phi}{dt^{j+k-2}}\right)_{1\le j,k\le n},
\end{equation} 
and
\begin{equation} \label{pf9}
\phi(t)=\frac{\sinh(2\ga)}{\sinh(t+\ga)\sinh(t-\ga)}\,.
\end{equation}

The aspect of the Izergin-Korepin determinantal formula that we exploit in this paper is that
$\tau_N$ can be expressed in terms of related orthogonal polynomials, see
 the paper  \cite{Z-J1} of Zinn-Justin.  In the ferroelectric phase the
expression in terms of orthogonal polynomials can be obtained in the following manner. 
For the evaluation of the Hankel determinant, let us
write $\phi(t)$ in the form of the Laplace transform of a discrete measure,
\begin{equation} \label{dph6}
\phi(t)=\frac{\sinh(2\ga)}{\sinh(t+\ga)\sinh(t-\ga)}=
4\sum_{l=1}^\infty e^{-2tl}\sinh(2\ga l).
\end{equation} 
Then
\begin{equation} \label{dph7}
\tau_n=\frac{2^{n^2}}{n!}\sum_{l_1,\ldots,l_n=1}^\infty \Delta(l_i)^2\prod_{i=1}^n
\left[2e^{-2tl_i}\sinh(2\ga l_i)\right], 
\end{equation}
where
\begin{equation} \label{dph8}
\Delta(l_i)=\prod_{1\le i<j\le n}(l_j-l_i)
\end{equation}
is the Vandermonde determinant. 

Introduce now discrete monic polynomials $P_j(x)=x^j+\dots$ orthogonal 
on the set
$\N=\{l=1,2,\ldots\}$ with respect to the weight,
\begin{equation} \label{dph9}
w(l)=2e^{-2tl}\sinh(2\ga l)=e^{-2tl+2\ga l}-e^{-2tl-2\ga l},
\end{equation}
so that
\begin{equation} \label{dph14}
\sum_{l=1}^\infty P_j(l)P_k(l)w(l)=h_k\delta_{jk}.
\end{equation}
Then it follows from (\ref{dph7}) that
\begin{equation} \label{dph15}
\tau_n=2^{n^2}\prod_{k=0}^{n-1}h_k, 
\end{equation}
see Appendix in the end of the paper \cite{BL}.

\subsection{Exact solution of $Z_n$ for finite $n$ on the critical line between the ferroelectric and disordered phases} 

We consider the partition  function $Z_n$  on the critical line
\begin{equation} \label{cr1}
\frac{b}{c}-\frac{a}{c}=1.
\end{equation}
We fix a point,
\begin{equation} \label{cr2}
\frac{a}{c}=\frac{\al-1}{2}\,,\quad \frac{b}{c}=\frac{\al+1}{2}\,;\qquad \al>1,
\end{equation}
on this line, and we are interested in the large $n$ asymptotics of the partition
function 
\begin{equation} \label{cr3}
Z_n=Z_n\left(\frac{\al-1}{2}\,,\frac{\al-1}{2}\,,\frac{\al+1}{2}\,,\frac{\al+1}{2}\,,1,1\right).
\end{equation}
Let us first derive a formula for $Z_n$ on the critical line. To that end, consider
the limit in the Izergin-Korepin formula in the ferroelectric phase, (\ref{pf7}), as
\begin{equation} \label{cr4}
t,\ga\to +0,\qquad \frac{t}{\ga}\to \al.
\end{equation}
Observe that in this limit,
\begin{equation} \label{cr5}
\frac{a}{c}=\frac{\sinh(t-\ga)}{\sinh(2\ga)}\to \frac{\al-1}{2},\qquad 
\frac{b}{c}=\frac{\sinh(t+\ga)}{\sinh(2\ga)}\to \frac{\al+1}{2}.
\end{equation}
By (\ref{cl_17}), 
\begin{equation} \label{cr6a}
Z_n\left(\frac{a}{c},\frac{a}{c},\frac{b}{c},\frac{b}{c},1,1\right)=\frac{Z_n(a,a,b,b,c,c)}{c^{n^2}}\,,
\end{equation}
hence by (\ref{pf7}), and (\ref{dph15}),
\begin{equation} \label{cr6}
Z_n\left(\frac{a}{c},\frac{a}{c},\frac{b}{c},\frac{b}{c},1,1\right)=
\left[\frac{2\sinh(t-\ga)\sinh(t+\ga)}{\sinh(2\ga)}\right]^{n^2}\prod_{k=0}^{n-1}\frac{h_k}{(k!)^2}\,. 
\end{equation}
To deal with limit (\ref{cr4}) we  need to rescale the orthogonal polynomials $P_k(l)$. 
Introduce the rescaled variable,
\begin{equation} \label{cr7}
x=2tl-2\ga l,
\end{equation}
and the rescaled limiting weight,
\begin{equation} \label{cr8}
w_{\al}(x)=\lim_{t,\ga\to +0,\; \frac{t}{\ga}\to \al} (e^{-2tl+2\ga l}-e^{-2tl-2\ga l})=e^{-x}-e^{-rx},
\qquad r=\frac{\al+1}{\al-1}>1\,.
\end{equation}
Consider monic orthogonal polynomials $P_j(x;\al)$ satisfying the orthogonality condition,
\begin{equation} \label{cr9}
\int_0^\infty P_j(x;\al)P_k(x;\al)w_{\al}(x)dx=h_{k,\al}\de_{jk}\,.
\end{equation}
To find a relation between $P_k(l)$ and $P_k(x;\al)$, introduce the monic polynomials
\begin{equation} \label{cr9a}
\tilde P_k(x)=\De^kP_k(x/\De),
\end{equation}
where
\begin{equation} \label{cr10}
\De=2t-2\ga,
\end{equation}
and rewrite orthogonality
condition (\ref{dph14}) in the form 
\begin{equation} \label{cr11}
\sum_{l=1}^\infty \tilde P_j(l\De)\tilde P_k(l\De)w_\al(l\De)\De=\De^{2k+1}h_k\delta_{jk},
\end{equation}
which is a Riemann sum for the integral in orthogonality condition (\ref{cr9}). Therefore,
\begin{equation} \label{cr12}
\lim_{t,\ga\to +0,\; \frac{t}{\ga}\to \al}\tilde P_k(x)=P_k(x;\al),
\end{equation}
and
\begin{equation} \label{cr13}
\lim_{t,\ga\to +0,\; \frac{t}{\ga}\to \al}\De^{2k+1}h_k=h_{k,\al}.
\end{equation}
Let us rewrite formula (\ref{cr6}) as
\begin{equation} \label{cr14}
Z_n\left(\frac{a}{c},\frac{a}{c},\frac{b}{c},\frac{b}{c},1,1\right)=
\left[\frac{2\sinh(t-\ga)\sinh(t+\ga)}{\sinh(2\ga)\De}\right]^{n^2}\prod_{k=0}^{n-1}\frac{\De^{2k+1}h_k}{(k!)^2}\,, 
\end{equation}
and take limit (\ref{cr4}). In the limit we obtain that
\begin{equation} \label{cr15}
Z_n=Z_n\left(\frac{\al-1}{2}\,,\frac{\al-1}{2}\,,\frac{\al+1}{2}\,,\frac{\al+1}{2}\,,1,1\right)=
\left(\frac{\al+1}{2}\right)^{n^2}\prod_{k=0}^{n-1}\frac{h_{k,\al}}{(k!)^2}\,,.
\end{equation}

Our main technical result in this paper will be the proof of the following asymptotics of $h_{k,\al}$.
Let, as usual, 
\begin{equation} \label{cr16}
\zeta(s)=1+\frac{1}{2^s}+\frac{1}{3^s}+\ldots,\qquad \Re s>1.
\end{equation}

\begin{theo} \label{h} As $k\to\infty$,
\begin{equation} \label{cr17}
\ln\left[\frac{h_{k,\al}}{(k!)^2}\right]=-\frac{\zeta(\frac{3}{2})}{2 \sqrt{\pi (r-1)} k^{1/2}}
+\frac{1}{4k}+O(k^{-3/2}),\qquad r=\frac{\al+1}{\al-1}\,.
\end{equation}
\end{theo}

A proof of this theorem will be given below.

\subsection{Main result} This paper is a continuation of the works \cite{BF} and
\cite{BL}, in which the large $n$ asymptotics of $Z_n$ is obtained in the
disordered and ferroelectric phase, respectively. In \cite{BF} it is proven that
in the disordered phase, for some $\ep>0$, as $n\to\infty$,
\begin{equation} \label{main1}
Z_n\left(\frac{a}{c},\frac{a}{c},\frac{b}{c},\frac{b}{c},1,1\right)=Cn^\kappa F^{n^2}
[1+O(n^{-\ep})],
\end{equation}
where in parameterization (\ref{pf6}),
\begin{equation} \label{main2}
\kappa=\frac{1}{12}-\frac{2\ga^2}{3\pi(\pi-\ga)}\,,
\end{equation}
and
\begin{equation} \label{main3}
F=\frac{\pi\sin(\ga-t)\sin(\ga+t)}{2\ga\sin(2\ga)\cos \frac{\pi t}{2\ga}}\,.
\end{equation}
The value of the constant $C$ in (\ref{main1}) is not yet known. In \cite{BL} it is proven that
in the  ferroelectric phase, for any $\ep>0$, as $n\to\infty$,
\begin{equation} \label{main4}
Z_n\left(\frac{a}{c},\frac{a}{c},\frac{b}{c},\frac{b}{c},1,1\right)=C G^n F^{n^2}
\left[1+O\left(e^{-n^{1-\ep}}\right)\right],
\end{equation}
where in parameterization (\ref{pf4}),
\begin{equation} \label{main5}
C=1-e^{-4|\ga|},\qquad G=e^{|\ga|-t},
\end{equation}
and
\begin{equation} \label{main6}
F=\frac{\sinh(t+\ga)}{\sinh(2\ga)}\,.
\end{equation}

The main result of this paper is the following asymptotics of $Z_n$ on the critical line between these two phases.

\begin{theo}\label{main} As $n\to\infty$,
\begin{equation} \label{main7} 
Z_n\left(\frac{\al-1}{2}\,,\frac{\al-1}{2}\,,\frac{\al+1}{2}\,,\frac{\al+1}{2}\,,1,1\right)=C n^\kappa
G^{\sqrt n}F^{n^2}[1+O(n^{-1/2})]\,,
\end{equation}
where $C>0$,
\begin{equation} \label{main8}
\kappa=\frac{1}{4}\,,\qquad G=\exp\left[-\zeta\left(\frac{3}{2}\right)\sqrt{\frac{\al-1}{ 2\pi }}\right]\,,
\end{equation}
and
\begin{equation} \label{main9}
F=\frac{\al+1}{2}\,.
\end{equation}
\end{theo}

The proof of Theorem \ref{main} follows easily from Theorem \ref{h}.
Namely, from formula (\ref{cr15}) and asymptotics (\ref{cr17}) we obtain that
\begin{equation} \label{main10}
\begin{aligned}
\ln &\left[\frac{Z_n\left(\frac{\al-1}{2}\,,\frac{\al-1}{2}\,,\frac{\al+1}{2}\,,\frac{\al+1}{2}\,,1,1\right)}
{\left(\frac{\al+1}{2}\right)^{n^2}}\right]
=\sum_{k=0}^{n-1}\ln\left[\frac{h_{k,\al}}{(k!)^2}\right]\\
&=\sum_{k=0}^{n-1} \left[-\frac{\zeta(\frac{3}{2})}{2 \sqrt{\pi (r-1)}\, k^{1/2}}
+\frac{1}{4k}+O(k^{-3/2})\right]\\
&=-\zeta\left(\frac{3}{2}\right)\sqrt{\frac{(\al-1) }{ 2\pi }}\,n^{1/2}
+\frac{\ln n}{4}+C_0+O(n^{-1/2})\,,
\end{aligned}
\end{equation}
which implies Theorem \ref{main}.

\subsection{Ground state configuration on the critical line}\label{ground_state}

The ground state configuration $\sg^{\rm gs}$ has the maximal weight. On the
upper critical line between the ferroelectric and disordered phase regions we have that $b=a+c$.
We also assume  that $a>0$, $c>0$,
hence $\sg^{\rm gs}$ should contain as many $b$'s as possible.
The domain wall boundary conditions imply that in each row there is at least one $c$.
Therefore, any weight cannot be bigger than $b^{n^2-n}c^n$. The weight $b^{n^2-n}c^n$ does occur
for the following unique configuration:

%%%%%%%%%%%% Fig.  %%%%%%%%%%%%%%
\begin{center}
 \begin{figure}[h]\label{gs}
\begin{center}
   \scalebox{0.7}{\includegraphics{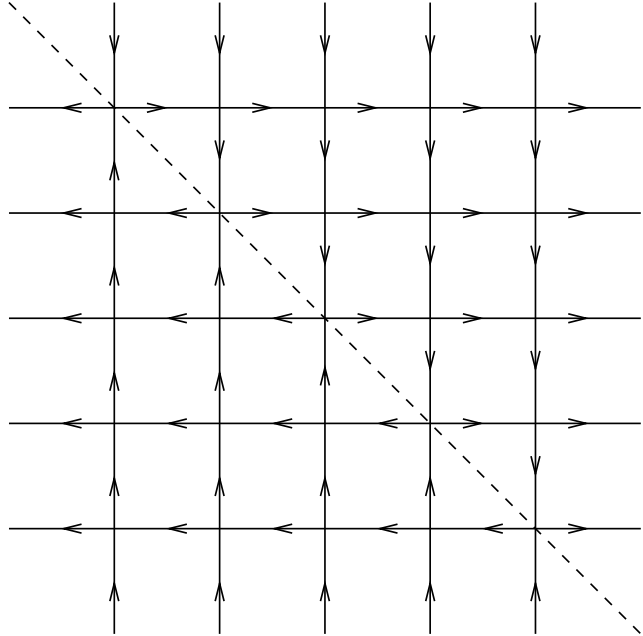}}
\end{center}
        \caption[A ground state configuration. ]
{A ground state configuration. }
    \end{figure}
\end{center}
%%%%%%%%%%%%%%%%%%%%%%%%%%%%%%%%%%

\begin{equation} \label{gs1}
\sg^{\rm gs}(x)=\left\{
\begin{aligned}
&\sg_5\quad{\rm if }\; x\;\text{\rm is on the diagonal},\\
&\sg_3\quad{\rm if }\; x\;\text{\rm is above the diagonal},\\
&\sg_4\quad{\rm if }\; x\;\text{\rm is below the diagonal},
\end{aligned}\right.
\end{equation}
see Fig.~4, which is the unique ground state configuration with domain wall boundary conditions
on the critical line. In the ferroelectric phase region, where $b>a+c$, the 
ground state configuration is obviously the same. If we set
\begin{equation} \label{gs2}
a=\frac{\al-1}{2}\,,\quad b=\frac{\al+1}{2}\,,\quad c=1,
\end{equation}
then the  weight of the ground state configuration is equal to
\begin{equation} \label{gs3}
w(\sg^{\rm gs})=F^{n^2-n},\qquad F=\frac{\al+1}{2}\,.
\end{equation}
By (\ref{main7}) the ratio $Z_n/w(\sg^{\rm gs})$ is evaluated as
\begin{equation} \label{gs4}
\frac{Z_n}{w(\sg^{\rm gs})}=Cn^\kappa G^{\sqrt n}F^n[1+O(n^{-1/2})].
\end{equation}
Observe that 
\begin{equation} \label{gs6}
\lim_{n\to\infty}\frac{\ln Z_n}{n^2}=\lim_{n\to\infty}\frac{\ln w(\sg^{\rm gs})}{n^2}=\ln F,
\end{equation}
so that the free energy is determined by the free energy
of the ground state configuration. This can be explained by the fact that
low energy excited states are local perturbations of the ground state around the diagonal. Namely, it
is impossible to create a new configuration by perturbing the ground state locally away of the diagonal:
the conservation law $N_3(\sg)=N_4(\sg)$ forbids such a configuration. Therefore, typical configurations
of the six-vertex model in the ferroelectric phase region and on the critical line
between the ferroelectric and disordered phase regions are frozen outside of a relatively
small neighborhood of the diagonal. 

This behavior of typical configurations in
the ferroelectric phase region  and on the critical line
between the ferroelectric and disordered phase regions is in a big contrast with
the situation in the disordered and anti-ferroelectric phase regions. Extensive rigorous, theoretical
and numerical studies, see, e.g., the works of  Cohn, Elkies, Propp \cite{CEP},
Eloranta \cite{Elo}, 
Syljuasen,  Zvonarev  \cite{SZ}, Allison, Reshetikhin \cite{AR}, Kenyon, Okounkov \cite{KO2},
Kenyon, Okounkov, Sheffield \cite{KOS}, Sheffield \cite{She},
Ferrari, Spohn \cite{FS}, Colomo, Pronko \cite{CP3},
Zinn-Justin \cite{Z-J2},
and references therein, show that in the disordered and anti-ferroelectric phase regions the
``arctic circle'' phenomenon persists, so that there are macroscopically big
frozen and random domains    in typical configurations, separated in the limit $n\to\infty$
by an ``arctic curve''.

\section{Large $k$ asymptotics of $h_{k,\al}$}

We will use
asymptotic formulae for orthogonal polynomials on $[0,\infty)$,
obtained in the paper \cite{Van} of Vanlessen. 
To formulate and to apply the Vanlessen's asymptotic formula we will need to introduce some notations
and to evaluate some parameters. 
Let us write
\begin{equation} \label{van1}
w_{\al}(z)=e^{-z}-e^{-rz}=ze^{-Q(z)}, 
\end{equation}
so that
\begin{equation} \label{van2}
Q(z)=z+\log \frac{z}{1-e^{-(r-1)z}}\,,
\end{equation}
where for $\log$ we take the principal branch with a cut at $(-\infty,0]$.
Observe that the function $Q(z)$ is analytic in a strip $|\Im z|\le c_0$, $c_0>0$.
Define the Mashkar-Rakhmanov-Saff numbers $\be_k=\be_k(\al)$ as a
solution to the equation
\begin{equation} \label{van3}
\frac{1}{2\pi}\int_0^{\be_k} Q'(x)\sqrt{\frac{x}{\be_k-x}}\,dx=k.
\end{equation}
As shown in \cite{Van}, for large $k$ there is a unique solution to this equation. 

\subsection{Evaluation of $\be_k$}
By the change of variable, $x=\be_k u$, equation (\ref{van3}) reduces to
\begin{equation} \label{van4}
\frac{\be_k}{2\pi}\int_0^1 Q'(\be_k u)\sqrt{\frac{u}{1-u}}\,du=k.
\end{equation}
Set
\begin{equation} \label{van5}
b_k=\frac{\be_k}{4k}\,.
\end{equation}
Then equation (\ref{van4}) reduces to
\begin{equation} \label{van6}
\frac{2 b_k}{\pi}\int_0^1 Q'(4 b_k k u)\sqrt{\frac{u}{1-u}}\,du=1.
\end{equation}
From (\ref{van2}),
\begin{equation} \label{van7}
Q'(z)=1+\frac{1}{z}-\frac{r-1}{e^{(r-1)z}-1}\,.
\end{equation}
Observe that the function $Q'(z)$ has poles at the points
\begin{equation} \label{van7a}
z=\frac{2m\pi i}{r-1}\,,\quad m=\pm1,\pm2,\ldots
\end{equation}
After evaluating integrals of the first two terms of $Q'(4 b_k k u)$, equation (\ref{van6}) reads
\begin{equation} \label{van8}
b_k+\frac{1}{2 k}-\frac{2}{\pi}\int_0^1 \frac{(r-1)b_k}{e^{4 (r-1)b_k ku}-1}\sqrt{\frac{u}{1-u}}\,du=1.
\end{equation}
By the change of variable $x=ku$, it reduces to
\begin{equation} \label{van9}
b_k+\frac{1}{2 k}-\frac{2}{\pi k^{3/2}}\int_0^k \frac{(r-1) b_k}{e^{4 (r-1) b_k x}-1}\sqrt{\frac{x}{1-(x/k)}}\,dx=1.
\end{equation}
Set
\begin{equation} \label{van10}
\ep=\frac{1}{k^{1/2}},
\end{equation}
and consider the function,
\begin{equation} \label{van11}
f(b,\ep)=b+\frac{\ep^2}{2}-\frac{2 \ep^3}{\pi}\int_0^{1/\ep^2} \frac{(r-1)b}{e^{4 (r-1) b x}-1}
\sqrt{\frac{x}{1-\ep^2 x}}\,dx-1,\qquad \frac{1}{2}\le b\le 2.
\end{equation}
Observe that as $\ep\to 0$,
\begin{equation} \label{van12}
\begin{aligned}
\int_0^{1/\ep^2} \frac{(r-1)b}{e^{4 (r-1) b x}-1}
\sqrt{\frac{x}{1-\ep^2 x}}\,dx&=\int_0^{\infty} \frac{(r-1)b\sqrt{x}\,dx}{e^{4 (r-1) b x}-1}
+O(\ep^2)\\
&=\frac{\sqrt{\pi}\,\zeta(\frac{3}{2})}{16\sqrt {b(r-1)}}+O(\ep^2),
\end{aligned}
\end{equation}
hence
\begin{equation} \label{van13}
f(b,\ep)=(b-1)+\frac{\ep^2}{2 }-\frac{\ep^3
\zeta(\frac{3}{2})}{8\sqrt {\pi b(r-1)}}+O(\ep^5).
\end{equation}
It is easy to see that equation (\ref{van12}) can be differentiated
in $b$ infinitely many times, and hence the function $f(b,\ep)$ is $C^\infty$ in a
neighborhood of the point $b=1,\;\ep=0$.
In addition,
\begin{equation} \label{van14}
f(1,0)=0,\qquad \frac{\partial f(1,0)}{\partial b}=1.
\end{equation}
By the implicit function theorem, there is a $C^\infty$-solution $b(\ep)$ of the
equation $f(b,\ep)=0$. From (\ref{van13})  we obtain that 
\begin{equation} \label{van15}
b(\ep)=1-\frac{\ep^2}{2 }+\frac{\ep^3\zeta(\frac{3}{2})}{8 \sqrt{\pi (r-1)} }+O(\ep^5),
\qquad \ep\to 0.
\end{equation}
Since $b_k=b(k^{-1/2})$, this gives
\begin{equation} \label{van16}
b_k=1-\frac{1}{2 k}+\frac{\zeta(\frac{3}{2})}{8 \sqrt{\pi (r-1)} k^{3/2}}+O(k^{-5/2}),
\qquad k\to\infty.
\end{equation}
By (\ref{van5}),
\begin{equation} \label{van17}
\be_k=4k b_k.
\end{equation}

\subsection{Evaluation of the equilibrium measure}
Set
\begin{equation} \label{em1}
V_{k}(x)=\frac{1}{k}\, Q(\be_k x),
\end{equation}
and consider the following minimization problem:
\begin{equation} \label{em2}
E=\inf_\mu I(\mu),
\end{equation}
where
\begin{equation} \label{em3}
 I(\mu)=-\iint\log|x-y| \,d\mu(x)d\mu(y)+\int V_k(x)d\mu(x).
\end{equation}
and $\di\inf_\mu$ is taken over all probability measures on $[0,\infty)$.
There exists a unique minimizer, $\mu=\mu_{k}$, and it has the following properties:
\begin{enumerate}
  \item The support of $\mu_{k}$ is the interval $[0,1]$.
  \item The measure $\mu_{k}$ is absolutely continuous with respect to the Lebesgue measure.
  \item The density function of $\mu_{k}$ has the form,
\begin{equation} \label{em4}
\frac{d\mu_{k}(x)}{dx}\equiv \psi_k(x)=\frac{1}{2\pi}\sqrt{\frac{1-x}{x}}\,q_k(x),
\end{equation} 
where $q_k(x)$ is analytic and positive on $[0,1]$.
\end{enumerate}
The equilibrium measure $\mu_k$ is characterized by the Euler-Lagrange variational conditions: 
there exists $l_k\in  \R$ such that
\begin{equation} \label{em5}
\begin{aligned}
2\int_0^1 \log |x - y|d\mu_k(y) - V_k(x) - l_k
& = 0\quad  \textrm{for}\; x \in [0,1], \\
2\int_0^1 \log |x - y|d\mu_k(y) - V_k(x) - l_k
& \le 0\quad  \textrm{for}\; x \not\in [0,1].
\end{aligned}
\end{equation}
The function $q_k(z)$ in (\ref{em4}) is given by the formula,
\begin{equation} \label{em6}
q_k(z)=\frac{1}{2\pi i}\oint_{\Ga}\sqrt{\frac{y}{y-1}}\,\frac{V_k'(y)\,dy}{y-z}\,,\qquad z\in \textrm{Int}\,\Ga,
\end{equation} 
where $\sqrt{\frac{y}{y-1}}$ is taken on the principal branch, with cut on $[0,1]$,
and $\Ga$ is a positively oriented contour containing $[0,1]\cup\{z\}$ 
in its interior, with the additional condition, that the function $V_k'(y)$ is analytic
inside $\Ga$. By (\ref{em1}) and (\ref{van2}),
\begin{equation} \label{em6a}
V_k(z)=4b_k z+\frac{1}{k}\ln\frac{4b_k k z}{1-e^{-4(r-1)b_k k z}},
\end{equation}
hence
\begin{equation} \label{em7}
V'_k(z)=4b_k+\frac{1}{k z}-\frac{\ga_k}{e^{\ga_k kz}-1}\,,
\end{equation}
where
\begin{equation} \label{em7a}
\ga_k=4(r-1)b_k,
\end{equation}
hence
\begin{equation} \label{em8}
q_k(z)=\frac{1}{2\pi i}\oint_{\Ga}\sqrt{\frac{y}{y-1}}\,
\left[4b_k+\frac{1}{k y}-\frac{\ga_k}{e^{\ga_k ky}-1}\right]\frac{dy}{y-z}\,.
\end{equation} 
By taking the residue at infinity, we obtain that
\begin{equation} \label{em9}
q_k(z)=4b_k+s_k(z),\qquad s_k(z)=-\frac{\ga_k}{2\pi i}\oint_{\Ga}\sqrt{\frac{y}{y-1}}\,
\frac{dy}{(e^{\ga_k ky}-1)(y-z)}\,.
\end{equation}
Observe that the function $V'_k(z)$ has poles at the points
\begin{equation} \label{em9b}
z_n=\frac{2\pi n i}{\ga_k k}\,,\quad n=\pm1,\pm2,\dots,
\end{equation}
hence the contour $\Ga$ has to pass close to 0. We choose $\Ga$ such that 
\begin{equation} \label{em9c}
\frac{c_1}{k}\ge\dist(0,\Ga)\ge \frac{c_2}{k},\quad c_1\ge \dist(1,\Ga)\ge c_2>0,
\end{equation}
see Fig.~5. More precisely, let for a given $z\in\C$, $m(z)\in[0,1]$ be the closest point from $z$
on $[0,1]$, so that
\begin{equation} \label{emz1}
\inf\{|z-u|,\; u\in[0,1]\}=|z-m(z)|.
\end{equation}
%%%%%%%%%%%% Fig.  %%%%%%%%%%%%%%
\begin{center}
\begin{figure}[h]\label{contour_Gamma}
\begin{center}
   \scalebox{0.5}{\includegraphics{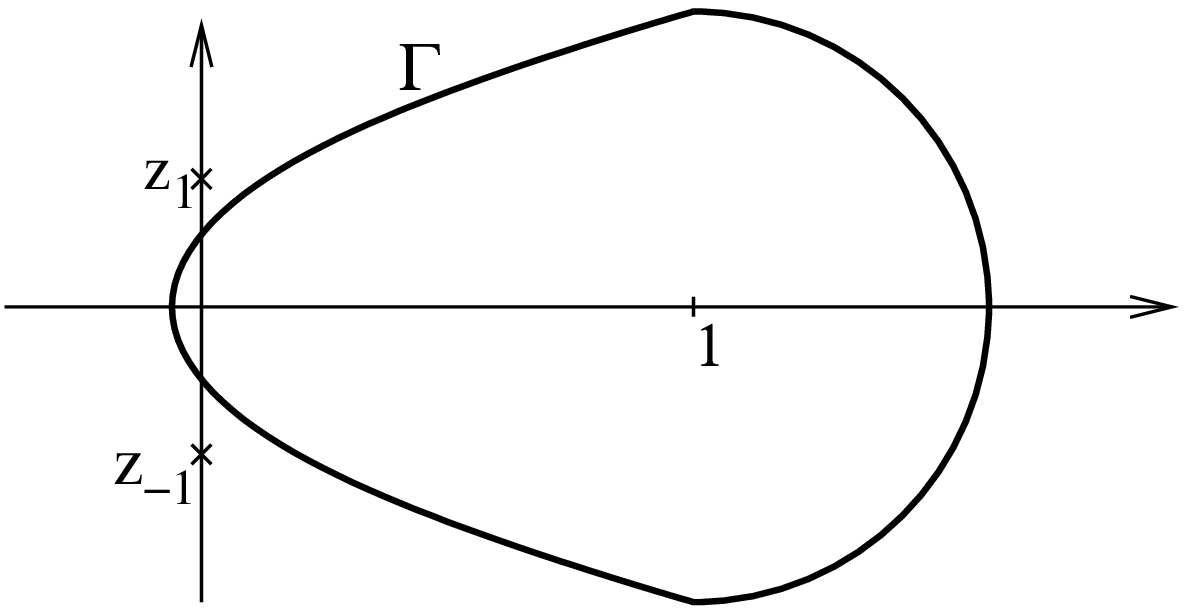}}
\end{center}
        \caption[The contour $\Gamma$.]{The contour $\Gamma$.}
    \end{figure}
\end{center}
%%%%%%%%%%%%%%%%%%%%%%%%%%%%%%%%%%
Then we define, for a given $\de>0$,
\begin{equation} \label{emz2}
\Ga=\Ga(\de,k)=\{ z\in\C:\; |z-m(z)|=\de\left[\frac{1}{k}+m(z)\right],
\end{equation}
and we choose $\de$ to be sufficiently small so that the points $z_n$ in (\ref{em9b})
lie outside of $\Ga$, see Fig.~5.
Observe that $\Ga(0,k)=[0,1]$.

With the help of the change of variables, $u=ky$, we obtain that
\begin{equation} \label{em9a}
s_k(z)=-\frac{\ga_k}{2\pi ik^{1/2}}\oint_{k\Ga}\sqrt{\frac{u}{(u/k)-1}}\,
\frac{du}{(e^{\ga_k u}-1)(u-kz)}\,,
\end{equation}
which implies that
\begin{equation} \label{em9d}
\sup_{0\le z\le 1} |s_k(z)|=O(k^{-1/2}),
\end{equation}
or even that
\begin{equation} \label{em9e}
\sup_{0\le d\le \frac{\de}{2}}\sup_{z\in\Ga(d,k)} |s_k(z)|=O(k^{-1/2}).
\end{equation}
For $z>1$, the function $s_k(z)$ can be reduced to
\begin{equation} \label{em10}
\begin{aligned}
s_k(z)&=-\ga_k\sqrt{\frac{z}{z-1}}\,\frac{1}{(e^{\ga_k kz}-1)}-\frac{\ga_k}{\pi }\int_0^1\sqrt{\frac{y}{1-y}}\,
\frac{dy}{(e^{\ga_k ky}-1)(z-y)}\,.
\end{aligned}
\end{equation}
It implies that
\begin{equation} \label{em11}
s_k(z)=\frac{a_k}{z}+r_k(z)\,,
\end{equation}
where
\begin{equation} \label{em12}
a_k=-\frac{\ga_k}{\pi}\int_0^1\sqrt{\frac{y}{1-y}}\,
\frac{dy}{(e^{\ga_k k y}-1)}\,
\end{equation}
and $r_k(z)$ satisfies the estimate,
\begin{equation} \label{em13}
|r_k(z)|\le \frac{C}{z\sqrt{z-1}\,k^{5/2}}\,,\quad z>1,
\end{equation}
with some $C>0$.
Indeed, 
\begin{equation} \label{em14}
\begin{aligned}
r_k(z)&=-\ga_k\sqrt{\frac{z}{z-1}}\,\frac{1}{e^{\ga_k kz}-1}-\frac{\ga_k}{\pi }\int_0^1\sqrt{\frac{y}{1-y}}\,
\frac{ydy}{(e^{\ga_k ky}-1)(z-y)z}\,.
\end{aligned}
\end{equation}
The first term on the right is exponentially small in $z$ and $k$, and it obviously satisfies
estimate (\ref{em13}). In the second term on the right, let us split the integral in two integrals,
from 0 to $\frac{1}{2}$ and from $\frac{1}{2}$ to 1. The first part is estimated as follows:
\begin{equation} \label{em15}
\begin{aligned}
0&\le \int_0^{\frac{1}{2}}\sqrt{\frac{y}{1-y}}\,
\frac{ydy}{(e^{\ga_k ky}-1)(z-y)z}\le \frac{4}{z^2}\int_0^{\frac{1}{2}}
\frac{y^{3/2}dy}{(e^{\ga_k ky}-1)}\\
&\le \frac{4}{z^2 k^{5/2}}\int_0^{\infty}
\frac{u^{3/2}du}{(e^{\ga_k u}-1)}\le \frac{C_0}{z^2 k^{5/2}}\,,\quad u=ky,
\end{aligned}
\end{equation}
hence it satisfies estimate (\ref{em13}). For the second part we have that
\begin{equation} \label{em16}
\begin{aligned}
0&\le \int_{\frac{1}{2}}^1\sqrt{\frac{y}{1-y}}\,
\frac{ydy}{(e^{\ga_k ky}-1)(z-y)z}\le \frac{1}{z(e^{\ga_k k/2}-1)}\int_{\frac{1}{2}}^1
\frac{dy}{(z-y)\sqrt{1-y}}\\
&= \frac{1}{z(e^{\ga_k k/2}-1)}\int_0^{\frac{1}{2}}
\frac{du}{(z-1+u)\sqrt{u}}\le \frac{C_1}{z\sqrt{z-1}\,(e^{\ga_k k/2}-1)}\,,\quad u=1-y,
\end{aligned}
\end{equation}
which satisfies estimate (\ref{em13}). Thus, (\ref{em13})  is proved.

From (\ref{em11}) we obtain that
\begin{equation} \label{em17}
q_k(z)=4b_k+\frac{a_k}{z}+r_k(z)\,,
\end{equation}
where $a_k$ is given by formula (\ref{em12}) and $r_k(z)$ satisfies estimate (\ref{em13}).

\subsection{Evaluation of the Lagrange multiplier} Introduce the function
\begin{equation} \label{lm1}
g_k(z)=\int_0^1 \log (z-x) d\mu_k(x),\quad z\in\C\setminus[0,1],
\end{equation}
where the branch of $\log$ is taken on the principal sheet, with a cut on $(-\infty,0]$. 
Then
\begin{equation} \label{lm2}
\om_k(z)\equiv g'_k(z)=\int_0^1 \frac{d\mu_k(x)}{z-x},\quad z\in\C\setminus[0,1].
\end{equation}
From equation (\ref{lm1}) it follows that as $z\to\infty$,
\begin{equation} \label{lm3}
g_k(z)=\log z+O(z^{-1}),
\end{equation}
and from (\ref{lm2}), that
\begin{equation} \label{lm4}
\om_k(z)=\frac{1}{z}+O(z^{-2}).
\end{equation}
From equation (\ref{em5}) it follows that
\begin{equation} \label{lm5}
\om_k(z)=\frac{V'_k(z)}{2}-\sqrt{\frac{z-1}{z}}\,\frac{q_k(z)}{2}\,,
\end{equation}
see, e.g., equations (3.27), (3.29) in \cite{Van}, and 
\begin{equation} \label{lm6}
l_k=2g_k(1)-V_k(1).
\end{equation}
Since
\begin{equation} \label{lm7}
\begin{aligned}
-g_k(1)=\lim_{u\to\infty}[g_k(u)-\log u-g_k(1)]&=\lim_{u\to\infty}\int_1^u \left[\om_k(z)-\frac{1}{z}\right]dz\\
&=\int_1^\infty\left[\frac{V'_k(z)}{2}-\sqrt{\frac{z-1}{z}}\,\frac{q_k(z)}{2}-\frac{1}{z}\right]dz,
\end{aligned}
\end{equation}
we obtain that
\begin{equation} \label{lm8}
g_k(1)=-\int_1^\infty\left[\frac{V'_k(z)}{2}-\sqrt{\frac{z-1}{z}}\,\frac{q_k(z)}{2}-\frac{1}{z}\right]dz,
\end{equation}
hence
\begin{equation} \label{lm9}
l_k=-\int_1^\infty\left[V'_k(z)-\sqrt{\frac{z-1}{z}}\,q_k(z)-\frac{2}{z}\right]dz-V_k(1).
\end{equation}
We split the last integral as
\begin{equation} \label{lm10}
\begin{aligned}
\int_1^\infty &\left[V'_k(z)-\sqrt{\frac{z-1}{z}}\,q_k(z)-\frac{2}{z}\right]dz
=\int_1^\infty\left[V'_k(z)-4b_k-\frac{1}{kz}\right]dz\\
&-\int_1^\infty\left[\sqrt{\frac{z-1}{z}}\,q_k(z)+\frac{2}{z}-4b_k-\frac{1}{kz}\right]dz=I_1-I_2.
\end{aligned}
\end{equation}
From (\ref{em7}) we have that
\begin{equation} \label{lm11}
I_1=\int_1^\infty\left[V'_k(z)-4b_k-\frac{1}{kz}\right]dz=
-\int_1^\infty \frac{\ga_kdz}{e^{\ga_k k z}-1}=O(e^{-c_0k})\,,\quad c_0>0.
\end{equation}
Let us evaluate $I_2$.
By (\ref{em17}),
\begin{equation} \label{lm12}
\begin{aligned}
I_2&=
\int_1^\infty\left[\sqrt{\frac{z-1}{z}}\,4b_k+\sqrt{\frac{z-1}{z}}\,\frac{a_k}{z}
+\sqrt{\frac{z-1}{z}}\,r_k(z)+\frac{2}{z}-4b_k-\frac{1}{kz}\right]dz.
\end{aligned}
\end{equation}
Since
\begin{equation} \label{lm13}
\begin{aligned}
\int_1^\infty\left(\sqrt{\frac{z-1}{z}}-1+\frac{1}{2z}\right)dz=\frac{1}{2}-\ln 2,
\end{aligned}
\end{equation}
we obtain that
\begin{equation} \label{lm14}
\begin{aligned}
I_2=b_k(2-4\ln 2)+
\int_1^\infty\left[\sqrt{\frac{z-1}{z}}\,\frac{a_k}{z}
+\sqrt{\frac{z-1}{z}}\,r_k(z)-\frac{2b_k-2}{z}-\frac{1}{kz}\right]dz.
\end{aligned}
\end{equation}
From estimate (\ref{em13}) we obtain that
\begin{equation} \label{lm15}
\int_1^\infty \sqrt{\frac{z-1}{z}}\,r_k(z)dz=O(k^{-5/2}),
\end{equation}
hence
\begin{equation} \label{lm16}
I_2=b_k(2-4\ln 2)+
\int_1^\infty\left[\sqrt{\frac{z-1}{z}}\,\frac{a_k}{z}
-\frac{2b_k-2+\frac{1}{k}}{z}\right]dz+O(k^{-5/2}).
\end{equation}
From equation (\ref{lm5}) we have that
\begin{equation} \label{lm17}
\om_k(z)=\frac{V'_k(z)}{2}-\frac{1}{2}\sqrt{\frac{z-1}{z}}\,\left[4b_k+\frac{a_k}{z}+r_k(z)\right].
\end{equation}
By equating terms of the order of $\frac{1}{z}$ on both sides, we obtain that
\begin{equation} \label{lm18}
1=\frac{1}{2k}-\frac{a_k}{2}+b_k,
\end{equation}
hence
\begin{equation} \label{lm19}
a_k=2b_k-2+\frac{1}{k}\,.
\end{equation}
By substituting this expression into (\ref{lm16}) we obtain that
\begin{equation} \label{lm20}
I_2=b_k(2-4\ln 2)+\left(2b_k-2+\frac{1}{k}\right)
\int_1^\infty\left(\sqrt{\frac{z-1}{z}}-1\right)\frac{dz}{z}+O(k^{-5/2}).
\end{equation}
Since
\begin{equation} \label{lm21}
\int_1^\infty\left(\sqrt{\frac{z-1}{z}}-1\right)\frac{dz}{z}=2\ln 2-2,
\end{equation}
we obtain that
\begin{equation} \label{lm22}
\begin{aligned}
I_2&=b_k(2-4\ln 2)+\left(2b_k-2+\frac{1}{k}\right)(2\ln 2-2)+O(k^{-5/2})\\
&=-2b_k-4\ln 2+4+\frac{2\ln2-2}{k}+O(k^{-5/2})\\
&=2-4\ln 2+\frac{2\ln2-1}{k}-\frac{\zeta(\frac{3}{2})}{4 \sqrt{\pi (r-1)} k^{3/2}}+O(k^{-5/2}).
\end{aligned}
\end{equation}
By (\ref{lm9}), (\ref{lm10}),
\begin{equation} \label{lm23}
l_k=I_2-I_1-V_k(1),
\end{equation}
and by (\ref{em6a}) and (\ref{van16}),
\begin{equation} \label{lm24}
\begin{aligned}
V_k(1)&=4b_k+\frac{\ln(4 b_k k)}{k}+O(k^{-5/2})\\
&=4+\frac{\ln k}{k}+\frac{2\ln 2-2}{k}
+\frac{\zeta(\frac{3}{2})}{2 \sqrt{\pi (r-1)} k^{3/2}}-\frac{1}{2k^2}+O(k^{-5/2}),
\end{aligned}
\end{equation}
hence
\begin{equation} \label{lm25}
l_k=-2-4\ln 2-\frac{\ln k}{k}+\frac{1}{k}
-\frac{3\zeta(\frac{3}{2})}{4 \sqrt{\pi (r-1)} k^{3/2}}+\frac{1}{2k^2}+O(k^{-5/2}),
\end{equation}

\subsection{Evaluation of $h_{k,\al}$} According to
Vanlessen's asymptotic formula, see \cite{Van},
\begin{equation} \label{hka1}
h_{k,\al}=\frac{\pi}{8}\,\be_k^{2k+2}e^{kl_k}\left[1+\left(\frac{3}{4q_k(0)}+\frac{47}{12q_k(1)}
-\frac{q_k'(1)}{4q_k(1)^2}\right)\frac{1}{k}+O(k^{-2})\right].
\end{equation}
Observe that in \cite{Van} this formula is proved under the assumption that the weight for orthogonal
polynomials has the form 
\begin{equation} \label{hka1a}
w(x)=x^{\ga}e^{-Q(x)},\qquad \ga>-1,
\end{equation}
where $Q(x)$ is a polynomial. In Appendix at the end of the paper we show what changes in the paper 
of Vanlessen \cite{Van} should be made to prove (\ref{hka1}) for the weight $w_\al(x)$ given by formula (\ref{van1}).
By (\ref{em9d}),
\begin{equation} \label{hka2}
q_k(0)=4+O(k^{-1/2}),\quad q_k(1)=4+O(k^{-1/2}).
\end{equation}
By (\ref{em9a}),
\begin{equation} \label{hka3}
q_k'(1)=-\frac{\ga_k}{2\pi ik^{1/2}}\oint_{k\Ga}\sqrt{\frac{u}{(u/k)-1}}\,
\frac{kdu}{(e^{\ga_k u}-1)(u-k)^2}\,,
\end{equation}
hence
\begin{equation} \label{hka4}
q_k'(1)=O(k^{-1/2}),
\end{equation}
because by condition (\ref{em9c}), the function $\frac{k}{u-k}$ is bounded by $1/c$
for $u\in k\Ga$.
From (\ref{hka2}) and (\ref{hka4}) we obtain that
\begin{equation} \label{hka5}
1+\left(\frac{3}{4q_k(0)}+\frac{47}{12q_k(1)}
-\frac{q_k'(1)}{4q_k(1)^2}\right)\frac{1}{k}=1+\frac{7}{6k}+O(k^{-3/2}).
\end{equation}
By substituting formulae (\ref{van17}), (\ref{van16}), (\ref{lm25}), and (\ref{hka5})
into (\ref{hka1}) and by using the Stirling formula for $k!$, we obtain that
\begin{equation} \label{hka6}
\ln\frac{h_{k,\al}}{(k!)^2}=-\frac{\zeta(\frac{3}{2})}{2 \sqrt{\pi (r-1)} k^{1/2}}
+\frac{1}{4k}+O(k^{-3/2})
\end{equation}
(we use MAPLE for this calculation). Theorem \ref{h} is proved.

\appendix

\section{Proof of formula (\ref{hka1})}

We use the notations and results from the work \cite{Van} of Vanlessen.  The essential difference
with \cite{Van} is that we consider not a fixed but a shrinking neighborhood of the origin, 
\begin{equation} \label{app1}
\tilde U_{\de,k}= \left\{z\in\C:\; |z|\le  \frac{\de}{k}\right\},
\end{equation}
where $\de>0$ is small enough so that the function $V_k(x)$ is analytic in $\tilde U_{\de,k}$,
see (\ref{em1}).
As in \cite{Van}, we consider a sequence of transformations of the Riemann-Hilbert problem
for orthogonal polynomials, and in the end we arrive at the following  Riemann-Hilbert problem 
on a $2\times 2$ matrix-valued function $R(z)$:
%%%%%%%%%%%% Fig.  %%%%%%%%%%%%%%
\begin{center}
\begin{figure}[h]\label{contour_Gamma_R}
\begin{center}
   \scalebox{0.5}{\includegraphics{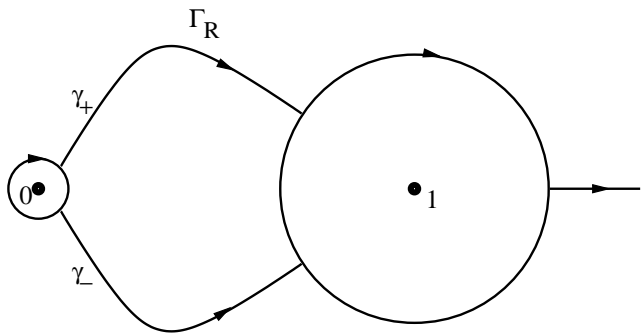}}
\end{center}
        \caption[The contour $\Gamma_R$.]{The contour $\Gamma_R$.}
    \end{figure}
\end{center}
%%%%%%%%%%%%%%%%%%%%%%%%%%%%%%%%%%
\begin{enumerate}
  \item $R(z)$ is analytic on $\C\setminus\Ga_R$, where $\Ga_R$ is the contour shown on Fig.~6, and it has limits,
  $R_+(z)$ and $R_-(z)$, on $\Ga_R$, as $z$ approaches a point on $\Ga_R$ from the left and from the right 
of the contour, with respect to the orientation indicated on Fig.~6.
  \item On $\Ga_R$, $R(z)$ satisfies the jump condition, $R_+(z)=R_-(z)v_R(z)$, where $v_R(z)$ is an explicit
 matrix-valued function.
  \item $R(z)\simeq I+\frac{R_1}{z}+\ldots$ as $z\to\infty$, where $I$ is the identity matrix.
\end{enumerate}
The contour $\Ga_R$ consists of the circle $\partial\tilde U_{\de,k}$, the circle $\partial U_{\de}$,
where
\begin{equation} \label{app1a}
U_{\de}= \left\{z\in\C:\; |z-1|\le  \de\right\},
\end{equation}
the boundaries of the lenses, $\ga_{\pm}$, and the semi-infinite interval $[1+\de,\infty)$.
The jump matrix $v_R$ on $\partial\tilde U_{\de,k}$ has the following asymptotics as $k\to\infty$:
\begin{equation} \label{app2}
v_R(z)\simeq I+\sum_{n=1}^\infty \tilde\Delta_n(z)k^{-n},
\end{equation}
see formulae (3.105) and (3.98) in \cite{Van}. This asymptotic formula holds under the condition
that $k^2z\to\infty$. Under this condition, for any $N\ge 1$ there exists a constant
$C_N>0$ such that
\begin{equation} \label{app2a}
\left|v_R(z)- I-\sum_{n=1}^N \tilde\Delta_n(z)k^{-n}\right|
\le\frac{C_N }{(k^2| z|)^{\frac{N}{2}}|z|^2}.
\end{equation}
The condition $k^2z\to\infty$ is valid for $z\in \partial\tilde U_{\de,k}$, and in this case 
the last estimate gives that
\begin{equation} \label{app2b}
\sup_{z\in \partial\tilde U_{\de,k}}\left|v_R(z)- I-\sum_{n=1}^N \tilde\Delta_n(z)k^{-n}\right|
\le\frac{\tilde C_N}{k^{\frac{N}{2}-2}},\qquad \tilde C_N=\frac{C_N}{\de^{\left[\frac{N}{2}\right]+2}}\,.
\end{equation}
The coefficients $\tilde\Delta_n(z)$ in (\ref{app2}) are given by the following formula:
\begin{equation} \label{app3}
\tilde\Delta_n(z)=\frac{1}{\tilde\phi_k(z)^{n/2}}P^{(\infty)}(z)
(-z)^{\frac{1}{2}\sg_3}A_n \,(-z)^{-\frac{1}{2}\sg_3}P^{(\infty)}(z)^{-1},
\end{equation}
where
\begin{align} 
\label{app4}
P^{(\infty)}(z)&=2^{-\sg_3}
\begin{pmatrix}
\frac{a(z)+a(z)^{-1}}{2} & \frac{a(z)-a(z)^{-1}}{2i} \\
\frac{a(z)-a(z)^{-1}}{-2i} & \frac{a(z)+a(z)^{-1}}{2} 
\end{pmatrix}\left(\frac{2z-1+2\sqrt {z(z-1)}}{z}\right)^{\frac{1}{2}\sg_3},\\ 
a(z)&=\left(\frac{z-1}{z}\right)^{1/4},\qquad
\label{app5}
\tilde\phi_k(z)=-\left[\frac{1}{4}\int_0^z\sqrt{\frac{1-s}{s}}\,q_k(s)\,ds\right]^2,\\
\label{app6}
A_n&=\frac{\prod_{j=1}^n [4-(2j-1)^2]}{16^{n} n!}
\begin{pmatrix}
\frac{(-1)^n}{4n}(3+2n) & (n-\frac{1}{2})i \\
(-1)^{n+1}(n-\frac{1}{2})i & \frac{1}{4n}(3+2n)
\end{pmatrix}, 
\end{align}
see equation (3.99) in \cite{Van}.
The function $\tilde\phi_k(z)$ is analytic in $\tilde U_{\de,k}$. From (\ref{em9}), (\ref{van16}),
and (\ref{em9e}) we obtain that as $k\to\infty$,
\begin{equation} \label{app7}
\sup_{z\in \tilde U_{\de,k}}|q_k(z)-4|=O(k^{-1/2}).
\end{equation}
By (\ref{app5}) this implies that
\begin{equation} \label{app8}
\sup_{z\in \tilde U_{\de,k}}\left|\frac{\tilde\phi_k(z)}{\phi(z)}-1\right|=O(k^{-1/2}),\qquad
\phi(z)=-\left(\int_0^z\sqrt{\frac{1-s}{s}}\,ds\right)^2=-4z+\frac{4z^2}{3}+\dots
\end{equation}
The function $\tilde\De_n(z)$ is meromorphic in $\tilde U_{\de,k}$ with the only
possible pole at the origin of the order at most $\left[\frac{n+1}{2}\right]$, see \cite{Van}.
This result, combined with explicit formula (\ref{app3}), implies that there exists $c_n>0$ such that
\begin{equation} \label{app8a}
\sup_{z\in \partial\tilde U_{\de,k}}\left|\tilde\Delta_n(z)k^{-n}\right|
\le c_n k^{-n+\left[\frac{n+1}{2}\right]}.
\end{equation}
This, in turn, allows us to improve estimate (\ref{app2b}) as follows: for any $N\ge 1$ 
there exists $\tilde c_N>0$ such that
\begin{equation} \label{app8b}
\sup_{z\in \partial\tilde U_{\de,k}}\left|v_R(z)- I-\sum_{n=1}^N \tilde\Delta_n(z)k^{-n}\right|
\le \tilde c_N k^{-N+\left[\frac{N}{2}\right]}\,,
\end{equation}
so that the error term is estimated by a constant times the estimate of $\tilde\Delta_{N+1}(z)k^{-(N+1)}$.
When $N=1$, this gives that
\begin{equation} \label{app9}
\sup_{z\in \partial\tilde U_{\de,k}}\left|v_R(z)-I- \frac{\tilde\Delta_1(z)}{k}\right|=O(k^{-1}).
\end{equation}
The function $\tilde\De_1(z)$ has a simple pole at 0 and its residue is equal to
\begin{equation} \label{app10}
B_k=\frac{3}{16q_k(0)}\,
2^{-\sg_3}
\begin{pmatrix}
1& i \\
i & -1
\end{pmatrix}2^{\sg_3}, 
\end{equation}
see equation (4.11) in \cite{Van}. The function $\tilde\Delta_1(z)-\frac{B_k}{z}$ is regular
at $z=0$ and from explicit formula (\ref{app3}) we obtain that as $k\to\infty$,  
\begin{equation} \label{app11}
\sup_{z\in \partial\tilde U_{\de,k}}\left|\tilde\Delta_1(z)-\frac{B_k}{z}\right|=O(1),
\end{equation}
hence from (\ref{app9}) we obtain that
\begin{equation} \label{app12}
\sup_{z\in \partial\tilde U_{\de,k}}\left|v_R(z)-I- \frac{B_k}{kz}\right|=O(k^{-1}).
\end{equation}
The problem here is that $v_R(z)$ is not close to $I$ on $\partial\tilde U_{\de,k}$.
We will overcome this obstacle by a transformation of the Riemann-Hilbert problem for $R(z)$.

Observe that 
\begin{equation} \label{app13}
\Tr B_k=0,\qquad \det B_k=0,
\end{equation}
hence
\begin{equation} \label{app14}
\det \left(I+\frac{B_k}{kz}\right)=1, 
\end{equation}
hence the matrix $I+\frac{B_k}{kz}$ is invertible for any $z\not=0$.
Let us make the substitution,
\begin{equation} \label{app15}
R(z)= 
\left\{
\begin{aligned}
 &\tilde R(z),\quad z\in \tilde U_{\de,k},\\
&\tilde R(z)\left(I+\frac{B_k}{kz}\right),\quad z\not\in \tilde U_{\de,k}.
\end{aligned}
\right.
\end{equation}
Then $\tilde R(z)$ solves the Riemann-Hilbert problem similar to the one for $R(z)$, with
the jump matrix $\tilde v_R(z)$ such that
\begin{equation} \label{app16}
\tilde v_R(z)=v_R(z)\left(I+\frac{B_k}{kz}\right)^{-1},\qquad z\in \partial\tilde U_{\de,k}
\end{equation}
and
\begin{equation} \label{app17}
\tilde v_R(z)=\left(I+\frac{B_k}{kz}\right)v_R(z)\left(I+\frac{B_k}{kz}\right)^{-1},
\qquad z\in \Ga_R\setminus\partial\tilde U_{\de,k}.
\end{equation}
From (\ref{app12}) we obtain that
\begin{equation} \label{app18}
\sup_{z\in \partial\tilde U_{\de,k}}\left|\tilde v_R(z)-I\right|=O(k^{-1}).
\end{equation}
Also, since the equilibrium density function diverges as $z^{-1/2}$ at the origin,
we obtain that $v_R(z)$ is sub-exponentially small on the boundary of lenses,
\begin{equation} \label{app19}
\sup_{z\in \ga_+\cup\ga_-}\left|v_R(z)-I\right|=O(e^{-c\sqrt{k}}),\qquad c>0.
\end{equation}
This implies that $\tilde v_R(z)$ satisfies a similar estimate,
\begin{equation} \label{app20}
\sup_{z\in \ga_+\cup\ga_-}\left|\tilde v_R(z)-I\right|=O(e^{-c\sqrt{k}}),\qquad c>0.
\end{equation}
In addition,
\begin{equation} \label{app21}
\sup_{z\in \partial U_{\de}}\left|\tilde v_R(z)-I\right|=O(k^{-1}),
\end{equation}
and
\begin{equation} \label{app22}
\left|\tilde v_R(z)-I\right|=O(e^{-ckz}),\quad z\ge 1; \qquad c>0.
\end{equation}
These estimates of smallness of $(\tilde v_R(z)-I)$ on $\Ga_R$ enable us to solve the
Riemann-Hilbert problem for $\tilde R(z)$ by a series of perturbation
theory. The fact that the radius of $\tilde U_{\de,k}$, $r=\frac{\de}{k}$, is tending 
to zero does not cause a problem, see appendix to the work \cite{BK}
of Bleher and Kuijlaars. 

The rest of the proof of formula (\ref{hka1}) goes along the lines
of \cite {Van}. Namely, by formula (4.17) in \cite{Van},
\begin{equation} \label{app23}
h_{k,\al}=\frac{\pi}{8}\,\be_k^{2k+2}e^{kl_k}\left[1-16i(R_1)_{12}+O(k^{-2})\right].
\end{equation}
By (\ref{app15}),
\begin{equation} \label{app24}
(R_1)_{12}
=(\tilde R_1)_{12}+\frac{(B_k)_{12}}{k}+O(k^{-2})
=(\tilde R_1)_{12}+\frac{3i}{64 q_k(0) k}+O(k^{-2})
\end{equation}
By applying formula (4.11) in \cite{Van} to $\tilde R_1$, we obtain that
\begin{equation} \label{app25}
(\tilde R_1)_{12}=-\frac{q_k'(1)i}{64 q_k(1)^2 k}+\frac{47i}{192 q_k(1) k}+O(k^{-2}).
\end{equation}
Observe that the first term in formula (4.11) in \cite{Van} is missing in this case,
because the function $\left[\tilde\Delta_1(z)-\frac{B_k}{z}\right]$ is regular at $z=0$. 
From the last two formulae we obtain that
\begin{equation} \label{app26}
-16i(R_1)_{12}=\left[\frac{3}{4 q_k(0) }-\frac{q_k'(1)}{4 q_k(1)^2}+\frac{47}{12 q_k(1)}\right]\frac{1}{k}+O(k^{-2}).
\end{equation}
By substituting this into (\ref{app23}) we obtain (\ref{hka1}).

\end{document}